\documentclass[twocolumn,epjd,superscriptaddress,noeprint]{revtex4}%
\usepackage{amsfonts}
\usepackage{amsmath}
\usepackage{amssymb}
\usepackage{graphicx}%
\begin{document}
\title{A protocol of potential advantage in the low frequency range to \\gravitational wave detection with space based optical atomic clocks}
\author{Feifan He}
\author{Baocheng Zhang}
\email{zhangbc.zhang@yahoo.com}
\affiliation{School of Mathematics and Physics, China University of Geosciences, Wuhan
430074, China}
\begin{abstract}
A recent proposal describes space based gravitational wave (GW) detection with
optical lattice atomic clocks [Kolkowitz et. al., Phys. Rev. D \textbf{94},
124043 (2016)] \cite{kpy16}. Based on their setup, we propose a new
measurement method for gravitational wave detection in low frequency with
optical lattice atomic clocks. In our method, n successive Doppler signals are
collected and the summation for all these signals is made to improve the
sensitivity of the low-frequency GW detection. In particular, the improvement
is adjustable by the number of Doppler signals, which is equivalent to that
the length between two atomic clocks is increased. Thus, the same sensitivity
can be reached but with shorter distance, even though the acceleration noises
lead to failing to achieve the anticipated improvement below the inflection
point of frequency which is determined by the quantum projection noise. Our
result is timely for the ongoing development of space-born observatories aimed
at studying physical and astrophysical effects associated with low-frequency GW.

\end{abstract}
\maketitle


\section{Introduction}

Direct detection of gravitational wave (GW) carries important implications
both for astronomy where information about astrophysical sources can be
obtained and for fundamental physics where aspects of relativistic theories of
gravity can be tested \cite{kst87}. In 2016, the first detection of GW from
two merging black holes \cite{bpa16} was reported by the advanced Laser
Interferometer Gravitational-wave Observatory (aLIGO), the famous terrestrial
laser interferometer observatory. A number of analogous GW events were
detected \cite{bpa162,bpa17,bpa172,bpa173} subsequently, all lying in the
frequency range of above dozens of Hertz (Hz). In the lower frequency range,
where prospective GW sources might stem from cosmological origin, such as the
very early phase of the Big Bang, or the more speculative astrophysical
objects like cosmic strings or domain boundaries \cite{aas12,aab12}, GW
remains elusive due to insufficient sensitivities. In fact, methods such as
laser interferometry in space \cite{lisa96,pas12,pas17,luo16,hw17}, pulsar
timing arrays \cite{dcb86,jcs10}, and Doppler tracking system \cite{jwa06}
etc., have been considered for detecting low frequency GW. Laser
interferometers with very long arm lengths can also be used to detect
low-frequency GW, although this remains as challenging. One of the earliest
and still popular proposals in this direction belongs to the Laser
Interferometer Space Antenna (LISA), whose planned launch date is arranged in
about 2034 \cite{lisa}. In view of the successful observations of GW, it is
important and timely to study other related methods for the detection of
low-frequency GW.

In this study, we investigate and extend the method of spacecraft Doppler
tracking. This precise technique traces back to the GP-A suborbital
experiments that measured the general relativistic redshift in the earth's
static gravitational fields \cite{rvc80}, although the facilitating idea that
fractional frequency fluctuation caused by GW on one-way Doppler of an
earth-based GW detector was studied already in 1970 \cite{wjf70}, and soon
afterwards, followed with a survey for its prospects of GW detection, by Davis
in two-way Doppler with deep space probes \cite{rwd74}. The traditional
technique for tracking distant spacecraft is to precisely monitor the Doppler
shift of a sinusoidal electromagnetic signal, which is continuously
transmitted to the spacecraft and coherently re-transmitted back to earth
\cite{jcb83}. In the Doppler tracking technique, the Earth and an
interplanetary spacecraft act as free test masses. The Doppler tracking system
continuously measures their relative dimensionless velocity, $\Delta
\upsilon/c=\Delta\nu/\nu_{0}$, where $\Delta\upsilon$ is the relative
velocity, $\Delta\nu$ is the associated Doppler frequency change, and $\nu
_{0}$ is the carrier frequency of the microwave link. A gravitational wave of
strain amplitude $h\left(  t\right)  $ propagating through the radio link
causes small perturbations in the Doppler time series of $\Delta\nu\left(
t\right)  /\nu_{0}$ \cite{jwa06,ewr75}. Recently, Kolkowitz et al \cite{kpy16}
proposes a space-based gravitational wave detector consisting of two spatially
separated, drag-free satellites sharing ultrastable optical laser light over a
single baseline and augmented by dynamical decoupling \cite{bzy13} for
improved sensitivity. In their method, atomic clocks (instead of atomic
interferometers) serve as GW sensors \cite{phk13,jmh16,cwy15}, and the
cumulative large-momentum-transfer experienced in atomic interferometry is
introduced into a system of clocks to enhance detection sensitivity
\cite{nct17}. A Doppler tracking system for GW detection via Double Optical
Clock in Space (DOCS) is also proposed \cite{sww18}, in which the frequency
range covers $10^{-4}$ Hz to $10^{-2}$ Hz with an overall estimated
sensitivity of $5\times10^{-19}$.

Although the last two proposals mentioned above can be sensitive to the low GW
frequency around $10^{-3}$ Hz, they need the arm length (distance between two
satellites) to be 1AU, which requires high optical power due to optical
diffraction. And for the same reason, the signal recycling cavity used in LIGO
is impossible for long baseline space-based optical interferometers. We find
that we can use the similar scheme to Ref. \cite{kpy16} but performed with the
recycling laser pulses in space to increase equivalently the arm length of
shorter space-based interferometer. It is helpful for some plans like TianQin
or Taiji Program in Space, in both of which the length of laser path is about
${\sim}10^{8}$ m \cite{luo16,hw17}. This paper is structured as follows.
First, we introduce the method of Doppler tracking for GW detection, while the
one-way clock interferometry is introduced in detail in second section. In
third section, we investigate the theory about sensitivity curves and compare
the one-way method with the two-way or two one-way methods schematically, from
which one can see the longer Ramsey precession time can increase the
sensitivity of the optical atomic clock detector. In fourth section, we
propose the recycling method which is equivalently to increase the effective
arm length, and discuss its limits. Finally, we conclude in the fifth section.

\begin{figure}[ptb]
\centering
\includegraphics[width=3.5in]{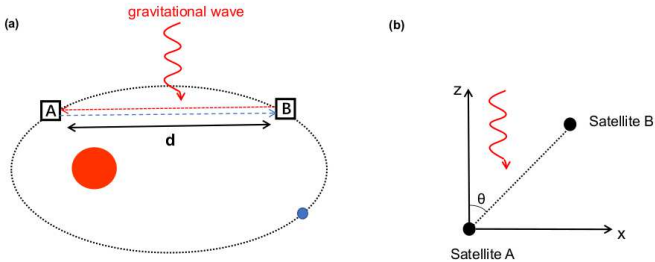} \caption{(Color online) (a) The
configuration for the proposed GW detector of Ref. \cite{kpy16}, which
consists of two identical drag-free satellites, A and B, in a heliocentric
orbit, separated from each other by a distance $d$. Each satellite maintains a
free-floating reference mass, an ultrastable laser, and a strontium optical
lattice clock. The measurement scheme we propose differs from that of Ref.
\cite{kpy16}, and will be discussed in Sect. III. (b) Illustration of the
geometric configuration.}%
\label{Fig1}%
\end{figure}

\section{Doppler Tracking Signal}

The scheme for Doppler tracking is shown in Fig. \ref{Fig1}(a), which is the
same as found in Ref. \cite{kpy16}, except that we will implement a different
operation protocol to be discussed in the fourth section. The two drag-free
satellites A and B are launched into a heliocentric orbit, each equipped with
an optical clock. The distance $d$ between the two satellites is set to
$5\times10^{10}$ m. Via two synchronized clocks and radio instruments on
board, a radio signal can be transmitted from A (B) to B (A), and the Doppler
signals as functions of time can be collected simultaneously on the two
satellites, i.e., two or bidirectional Doppler tracking measurements are
carried out simultaneously.

Figure \ref{Fig1}(b) illustrates the simple geometric configuration for the
discussed Doppler tracking system for GW detection, where satellite A is set
at the origin and the two satellites lie in the $x$-$z$ plane separated by a
distance $d$. Assuming that the GW is propagating along the $z$-axis direction
through the Doppler tracking system and the corresponding spacetime can be
described by the perturbed metric
\begin{equation}
ds^{2}=-c^{2}dt^{2}+\left(  1+h\right)  dx^{2}+\left(  1-h\right)
dy^{2}+dz^{2}, \label{gwm}%
\end{equation}
where $h=h\left(  t-z\right)  $ is exceedingly small compared with unity and
it describes the strain field of a train of plane gravitational waves. The
spacetime Eq. (\ref{gwm}) has symmetries generated by the Killing vectors:
\{$\frac{\partial}{\partial x},\frac{\partial}{\partial y},\frac{\partial
}{\partial z}+\frac{\partial}{\partial t}$\}.

The influence of GW on the signal of a Doppler tracking system is easily
calculated. For a light signal sent at time $t$ from system $A$ to system $B$,
the light at $A$ can be described by a null vector with the form%
\begin{equation}
\sigma_{0}=\left(  -\nu_{0}\right)  \left[  dt+\left(  1+h_{0}/{2}\right)
\sin\theta dx+\cos\theta dz\right]  ,
\end{equation}
with $\theta$ the angle between the link line $AB$ and $z$-axis. $\nu_{0}$ is
the observed frequency of the light, and $h_{0}$ is the value of $h\left(
t-z\right)  $ at the emitter $A$. When the light arrives at the receiver $B$,
its frequency and the GW strain becomes $\nu_{1}$ and $h_{1}$. Define the
frequency shift parameter
\begin{equation}
z\equiv\frac{\nu_{1}-\nu_{0}}{\nu_{0}}, \label{fsd}%
\end{equation}
for the situation considered here, it is given by \cite{ewr75},
\begin{equation}
z_{AB}=\frac{1}{2}\left(  1+\cos\theta\right)  \left(  h_{0}-h_{1}\right)  ,
\label{fss}%
\end{equation}
with $h_{1}=h\left[  t+\left(  1-\cos\theta\right)  {d}/{c}\right]  $
according to the coordinates used in Fig. \ref{Fig1}(b) where point $B$ is
specified by $x={d\sin\theta}$, $y=0$, and $z={d\cos\theta}$. The result given
in Eq. (\ref{fss}) is identical to that obtained from calculating directly the
change of the distance between $A$ and $B$ \cite{kpy16}. If $\theta=0$, the
light photons are sent out parallel to the GW normal, $z_{AB}=0$; If
$\theta={\pi}/{2}$, the photons intersect perpendicularly the direction of GW
propagation, $z_{AB}=\left[  h\left(  t\right)  -h\left(  t+{d}/{c}\right)
\right]  /2$. For simplicity, in what follows we take the latter case of
$\theta={\pi}/{2}$ to calculate the maximal GW signal. When a particular
Fourier component of the GW $h\left(  t\right)  =\left\vert h\right\vert
\sin\left(  2\pi ft+\varphi\right)  $ with an amplitude $\left\vert
h\right\vert $ and an arbitrary phase $\varphi$ is considered, the signal
becomes $z={\Delta\nu}/{\nu}=-\left\vert h\right\vert \cos\left[  2\pi
f\left(  t+{d}/{(2c)}\right)  +\varphi\right]  \sin(\pi f{d}/{c})$ for
perpendicular light propagation, which results in the maximal fractional
frequency difference between the two clocks occurring at $f=c/2d$.

At frequencies other than the optimal, the magnitude of the detectable GW
signal is modulated by the inherent sensitivity of the specific setup, as
captured by the detector's transfer function $\Gamma\left(  f\right)  $ and
the degree of system's susceptibility to noise \cite{crl03}. For the one-way
Doppler shift, Eq. (\ref{fss}) can be expressed in Fourier space as
\begin{equation}
Z_{AB}\left(  f\right)  =\frac{1}{2}H\left(  f\right)  \left(  1-e^{i2\pi
fd/c}\right)
\end{equation}
where capital lettered functions denote Fourier transforms $Z\left(  f\right)
=\int dte^{i2\pi ft}z\left(  t\right)  $ and $H\left(  f\right)  =\int
dte^{i2\pi ft}h\left(  t\right)  $. The modulation factor to $H(f)$,
$\Gamma_{\nu}\left(  f\right)  =\left\vert \left(  1-e^{i2\pi fd/c}\right)
/2\right\vert ^{2}=\sin^{2}\left(  \pi fd/c\right)  $, depends only on the
geometry of the detector and is called geometric transfer function, which
differs from its counterpart $\Gamma_{\phi}\left(  f\right)  =\sin$%
c$^{2}\left(  \pi fd/c\right)  $ for phase detectors \cite{rs97}.

The actual measured GW signal for the clock-based detector also depends on the
measurement scheme used for the atoms. A long integration time $T$ increases
the sensitivity, but is limited by atomic linewidth, $T_{\max}=1/(2\pi
\Delta_{A})\approx160$ s, where the transition linewidth is $\Delta_{A}=1$ mHz
\cite{lby15,bzl06}.\ The signal acquired for a clock measurement between
$t_{0}$ and $t_{0}+T$ is therefore given by
\begin{equation}
\bar{z}=\frac{1}{T}\left\vert \int_{t_{0}}^{T+t_{0}}z\left(  t\right)
dt\right\vert =\left\vert \int_{-\infty}^{\infty}dtF\left(  t_{0}-t\right)
z\left(  t\right)  \right\vert ,\label{aff}%
\end{equation}
where $F\left(  t\right)  $ describes a window function that captures the
measurement sequence of duration $T$ for a specific protocol. With the Ramsey
sequence ($\pi/2$ pulse at the beginning and another $\pi/2$ pulse at the end
and ignoring the finite pulse operation durations), the window function
reduces to $F\left(  t\right)  =1/T$ for $t\in\left[  -T,0\right]  $ and
$F\left(  t\right)  =0$ otherwise. For a continuous GW with $h\left(
t\right)  =\left\vert h\right\vert \sin\left(  2\pi ft+\varphi\right)  $, this
gives the one-way result as%
\begin{align}
\bar{z}_{AB} &  =\frac{\left\vert h\right\vert }{\pi fT}\left\vert \sin\left(
\pi f{d}/{c}\right)  \sin\left(  \pi fT\right)  \right.  \nonumber\\
&  \left.  \times\cos\left[  \pi f\left(  2t_{0}+{d}/{c}+T\right)
+\varphi\right]  \right\vert ,
\end{align}
The signal considered above is continuous. It can be made simpler by adapting
the starting time of the measurement to account for $\varphi$ and thus set the
argument of the cosine to $0$ to give the maximum%
\begin{equation}
\bar{z}_{AB}=\left\vert h\right\vert \left\vert \sin\left(  \pi f{d}%
/{c}\right)  \mathrm{sinc}\left(  \pi fT\right)  \right\vert .
\end{equation}

\section{Sensitivity Curve}

In the initial Doppler tracking scheme \cite{jwa06}, a two-way, or two
one-way-trip measurement method is used. A recent different proposal called
DOCS \cite{sww18}, on the other hand, makes use of the differential signal
from the two one-way-trip measurements. They are briefly summarized below and
compared to each other based on the protocol put forward in Ref. \cite{kpy16},
in search for any possible improvement for GW detection.

The two-way method involves light being reflected or re-emitted from $B$ along
the reverse trajectory (for the first trip of $A$ to $B$), and detected in the
end at the original place $A$ (at $t={2d}/{c}$). The reversed return trip for
the light is expressed as
\begin{equation}
\sigma_{0}^{^{\prime}}=\left(  -\nu_{1}\right)  \left(  dt-\left(  1+h_{1}%
/{2}\right)  \sin\theta dx-\cos\theta dz\right)  .
\end{equation}
When light returns to ${A}$, its frequency and strain are changed into
$\nu_{2}$ and $h_{2}=h\left(  t+{2d}/{c}\right)  $ respectively. Thus, an
observer at ${A}$ can calculate the Doppler shift of the returning light
\cite{ewr75} according to
\begin{align}
z_{ABA}=z_{AB}+z_{BA}  &  =\left(  h_{0}-h_{1}\right)  \left(  1+\cos
\theta\right)  /{2}\nonumber\\
&  +\left(  h_{1}-h_{2}\right)  \left(  1-\cos\theta\right)  /{2}%
.\ \ \ \ \ \ \label{fsdd}%
\end{align}
As in the case of the one-way method above, one can obtain the corresponding
two-way result as%
\begin{equation}
\bar{z}_{ABA}=\left\vert h\right\vert \left\vert \sin\left(  2\pi f{d}%
/{c}\right)  \mathrm{sinc}\left(  \pi fT\right)  \right\vert .
\end{equation}
where it is noticed that when the frequency $f=m/T$ ($m$ is the integer) and
distance $d=cT/2$, some zero points are taken, which will lead to the
divergence for the calculation of sensitivity below. However, the infinity is
not seen in all figures about the sensitivity curves because data point
acquisition is not dense enough in our numerical calculation.

Usually, one expects the signal to noise can be improved by the constraint of
the common noise modes when the signals of the two separate satellites are
compared, as discussed before \cite{mt96,prv86,lhh00,tzz15}. Based on the
scheme of Fig. \ref{Fig1}, an improved low-frequency result might be obtained
from the difference of the two one-way signals, as was studied recently in
Ref. \cite{sww18} and given by
\begin{equation}
\bar{z}_{ABD}=\left\vert h\right\vert \left\vert \left(  1-\cos\left(  2\pi
f{d}/{c}\right)  \right)  \mathrm{sinc}\left(  \pi fT\right)  \right\vert .
\end{equation}
where the average is made according to Eq. (\ref{aff}) with $z\left(
t\right)  $ replaced by the formula $z_{ABD}=z_{AB}-z_{BA}$. Although
experimental conditions like noises might be different for the two way and two
one-way implementation, it is concerned here that how the two data
combinations resist the quantum projection noise respectively through the
discussion about the sensitivity below.

\begin{figure}[ptb]
\centering
\includegraphics[width=3.35in]{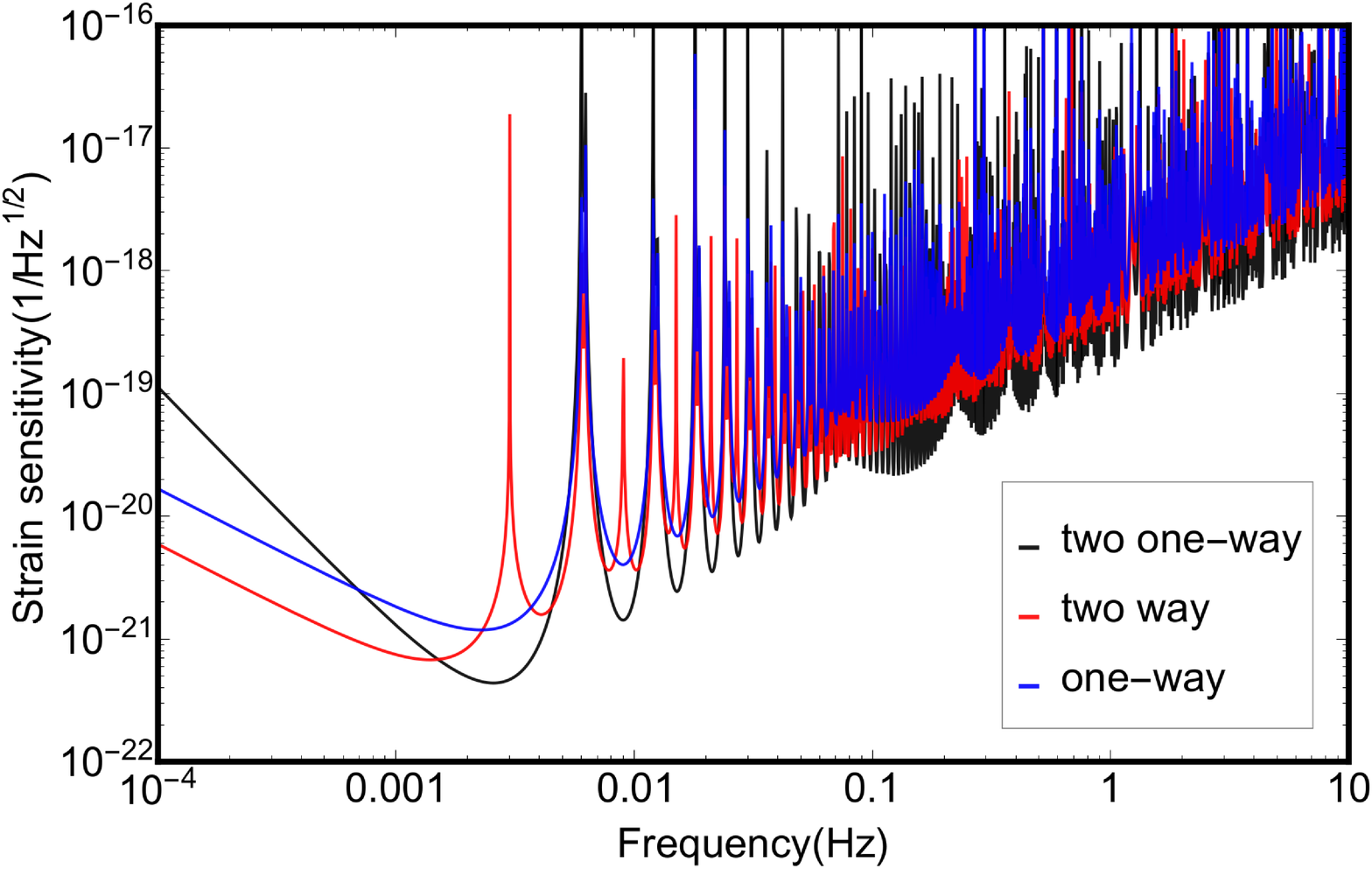} \caption{(Color online) Comparing the
sensitivity curves for one-way, two-way, and two one-way measurement
protocols.}%
\label{fig2}%
\end{figure}

In order to estimate the sensitivity, one analyzes the limit imposed by the
noise (as signal strength). Following the steps of Ref. \cite{kpy16}, the
smallest detectable fractional frequency difference for total measurement time
$\tau=1$ s constrained by noise is given by%
\begin{equation}
\sigma_{\min}\left(  \tau\right)  =\frac{\delta\nu_{\min}}{\nu}=\frac
{\sqrt{\Delta_{A}}}{\nu\sqrt{2\pi\tau N}}\simeq1.1\times10^{-20}%
/\sqrt{\mathrm{Hz}}\,,\label{sens}%
\end{equation}
where the frequency of the optical clock transition is $\nu=430$ THz
\cite{lby15,bzl06}, the transition linewidth is $\Delta_{A}=1$ mHz and the
atomic number is $N=7\times10^{6}$. Thus, the smallest measurable GW-induced
strain can be determined by $\bar{z}=\sigma_{\min}\left(  \tau\right)  $, but
the sensitivity curve is made using the general noise expression, $\sigma
^{2}\left(  \tau\right)  =\frac{1}{\left(  2\pi\upsilon\right)  ^{2}T\tau
}\left(  \frac{1}{N}+\frac{T}{\tau}\sqrt{\frac{h\upsilon\Delta_{L}}{\eta P}%
}\right)  $ where $\Delta_{L}$ is the linewidth of the laser, $P$ is the power
of the laser, and $\eta$ is the detector quantum efficiency, which is
presented in Fig. \ref{fig2}. It shows that two-way method can shift the
optimal measurement to a lower frequency, which can be extended to the general
case discussed in the next section.

\section{Recycling scheme}

The protocol we suggest consists of $n$ successive one-way-trip, or $n$-way
light propagation back and forth: the first laser light is sent at time $t$
from $A$ to $B$. The moment $B$ receives the light from $A$, it re-transmits a
laser to $A$. Such a sequence between $A$ and $B$ is alternately repeated with
every trip of the light path experiences the change due to GW. Noticed that in
the scheme described in Fig. \ref{Fig1}, the distance between two atomic
clocks just matches the measurement time determined by atomic linewidth, that
is $T_{\max}=1/(2\pi\Delta_{A})=160$ s. So here the shorter distance has to be
considered for the implementation of $n$-way method. For example, if the
distance $d=5\times10^{8}$ m is considered, $n$ will be constrained.

The signal for the $i$th one-way-trip is%
\begin{equation}
z_{i}=\frac{\nu_{i}-\nu_{i-1}}{\nu_{i-1}}=\frac{1}{2}\left(  1+\left(
-1\right)  ^{i-1}\cos\theta\right)  \left(  h_{i-1}-h_{i}\right)  ,
\end{equation}
where%
\begin{equation}
h_{i}=h\left(  t+\sum_{m=1}^{i}\left(  1+\left(  -1\right)  ^{m}\cos
\theta\right)  {d}/{c}\right)  .
\end{equation}

Summing up to get the total signal, we find%
\begin{align}
Z_{T}  &  =\frac{\nu_{n}-\nu_{0}}{\nu_{0}}=\sum_{m=1}^{n}z_{m}\left(  t\right)
\nonumber\\
&  =\frac{1}{2}\left(  1+\cos\theta\right)  h_{0}-\frac{1}{2}\left(  1+\left(
-1\right)  ^{n-1}\cos\theta\right)  h_{n}\nonumber\\
&  +\sum_{m=1}^{n-1}\left(  -1\right)  ^{m}\cos\theta h_{m},
\end{align}
which reduces to
\[
Z_{T}(t)=\frac{1}{2}h\left(  t\right)  -\frac{1}{2}h\left(  t+\frac{nd}%
{c}\right)  ,
\]
at $\theta={\pi}/{2}$ as before. It only contains signals from the first and
the last one-way-trips. Thus, the cumulative signal from our $n$-way
consecutive two point signals: $AB$, $BA$, $AB$, $\cdots$ sums up to an
effective one-way signal with the effective two point distance the total from
all one-way-trips $(\propto nd)$. This result resembles that of LIGO, a
Michelson interferometer, whose effective arm length is multiplied by
($\sim250$ times round trips) with a Fabry-Perot inserted into each arm. But
the LIGO arrangement cannot adapt to the space based setting we discuss, for
long baselines, the power received at satellite B is related to the power
transmitted from satellite A by $\ P_{B}=P_{A}(\pi R^{2}\nu/dc)^{2}$, where
$R$ is the telescope diameter on satellites A and B. For $R=20$ cm,
transmitted power of $P_{A}=1$ W and arm length $d=5\times10^{10}$ m, the
received optical power at satellite B would be $\sim300$ pW. If the light is
reflected to A from B, the power received at A will be $\sim10^{-7}$ pW, which
is too low to be detected. Our $n$-way protocol, however, overcomes such a
challenge by sending another laser pulse back after receiving instead of
reflecting by a mirror, this can be achieved by phase lock loop \cite{hgc96}.
Therefore, despite of the shrinking signal/noise from the diffractive loss,
the loss for each one-way-trip remains tolerable.

For a continuous GW with $h\left(  t\right)  =\left\vert h\right\vert
\sin\left(  2\pi ft+\varphi\right)  $, the above $Z_{T}$ reduces to%
\begin{align}
Z_{T}(t)  &  =\frac{\left\vert h\right\vert }{2}\left[  \sin\left(  2\pi
ft+\varphi\right)  -\sin\left(  2\pi f\left(  t+\frac{nd}{c}\right)
+\varphi\right)  \right] \nonumber\\
&  =\left\vert h\right\vert \cos\left(  2\pi f\left(  t+\frac{nd}{2c}\right)
+\varphi\right)  \sin\left(  \pi f\frac{nd}{c}\right)  .
\end{align}

\begin{figure}[ptb]
\centering
\includegraphics[width=3.35in]{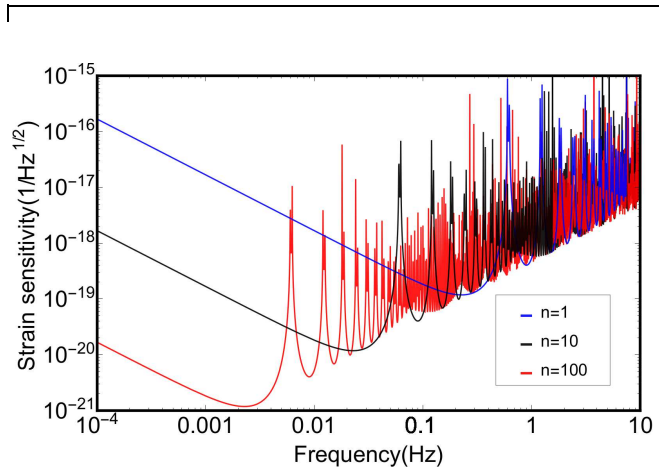} \caption{(Color online) Sensitivity
curves for $n=1$, $10$, and $100$-trip one-way laser propagation in the
presumed absence of acceleration noise.}%
\label{fig3}%
\end{figure}

\begin{figure}[ptb]
\centering
\includegraphics[width=3.35in]{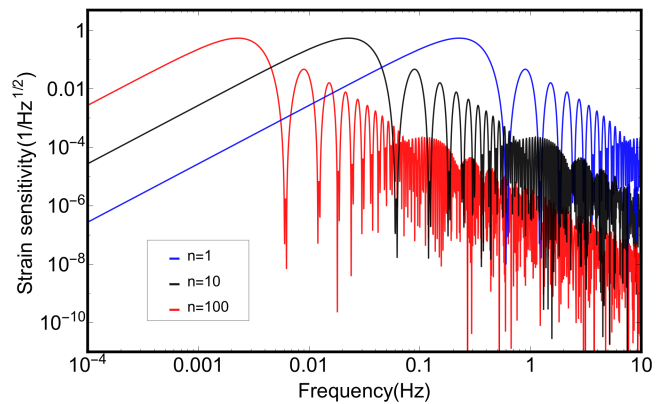} \caption{(Color online) The curves
for the transfer function with $n=1$, $10$, and $100$. }%
\label{fig4}%
\end{figure}

With a window function $F\left(  t\right)  =1/T$ for $t\in\left[  -T,0\right]
$ and $F\left(  t\right)  =0$ otherwise, the actual measured signal
becomes\begin{widetext}
\begin{align*}
\bar{Z}_{T}  &  =\frac{1}{T}\left\vert \int_{t_{0}}^{T+t_{0}}dtZ_{T}%
(t)\right\vert =\left\vert \int_{-\infty}^{\infty}dtF\left(  t_{0}-t\right)
Z_{T}(t)\right\vert \\
&  =\frac{\left\vert h\right\vert }{2\pi fT}\left\vert \left[  \sin\left(
2\pi f\left(  t_{0}+T+\frac{nd}{2c}\right)  +\varphi\right)  -\sin\left(  2\pi
f\left(  t_{0}+\frac{nd}{2c}\right)  +\varphi\right)  \right]  \sin\left(  \pi
f\frac{nd}{c}\right)  \right\vert \\
&  =\frac{\left\vert h\right\vert }{\pi fT}\cos\left(  2\pi f\left(
t_{0}+\frac{T}{2}+\frac{nd}{2c}\right)  +\varphi\right)  \sin\left(  \pi
fT\right)  \sin\left(  \pi f\frac{nd}{c}\right) \\
&  =\left\vert h\right\vert \cos\left(  2\pi f\left(  t_{0}+\frac{T}{2}%
+\frac{nd}{2c}\right)  +\varphi\right)  {\rm sinc}\left(  \pi fT\right)
\sin\left(  \pi f\frac{nd}{c}\right).
\end{align*}
\end{widetext}Adopting the measurement starting time $t_{0}$ to account for
$\varphi$ gives the maximum of the above
\begin{equation}
\bar{Z}_{T}=\left\vert h\right\vert \mathrm{sinc}\left(  \pi fT\right)
\sin\left(  \pi f\frac{nd}{c}\right)  , \label{nsm}%
\end{equation}
where the measurement time $T=nd/c$, and for $d=5\times10^{8}$ m, $n$ cannot
be larger than $100$ due to $T\leq T_{\max}=160$ s. The sensitivity curves
could be obtained by taking $\bar{Z}_{T}=\sigma\left(  \tau\right)  $ where
$\sigma\left(  \tau\right)  $ is the general noise expression given in the
discussion below Eq. (\ref{sens}), which are presented in Fig. \ref{fig3} for
$n=1$, $n=10$ and $n=100$ with $d=5\times10^{8}$ m. It is seen that when $n$
increases, the sensitivity is improved towards lower frequencies.
Quantitatively, we find as long as the GW frequency satisfies $c/2nd\leq f\leq
c/2d$, improved sensitivity can be expected. This implies that quantum
projection noise is constrained. From Eq. (\ref{nsm}), the transfer function
of our $n$-way protocol described above is given by%
\begin{equation}
\Gamma\left(  f\right)  =\mathrm{sinc}^{2}\left(  \pi fT\right)  \sin
^{2}\left(  \pi f{nd}/{c}\right)  ,
\end{equation}
which is graphed in Fig. \ref{fig4} for $n=1$, $n=10$ and $n=100$. It is seen
that this measurements yield the maximal signal for $f=c/2nd$. Moreover, we
check for the inclusion of a certain amount of dead time at points $A$ and
$B$, such that they could re-transmit after detecting the arrival of incoming
light. We find that the constraint on quantum projection is not as good as the
present method, but could also be improved when the $n$ increases.

However, at low frequencies, the main noise is derived from the residual
acceleration noise of the free reference masses, which causes the sensitivity
to scale as $1/f^{2}$ similar to that analyzed for LISA
\cite{lhh00,dpc11,aad12}. As an estimate, we take the spectral density of
phase noise contributed by acceleration noise as $S_{pa}\left(  f\right)
={S_{a}}/{\left(  2\pi f\right)  ^{4}\left(  c\tau\right)  ^{2}}$
\cite{lhh00}, where acceleration noise spectrum is at a level of
$S_{a}=9\times10^{-30}m^{2}s^{-4}\mathrm{Hz}^{-1}$.

\begin{figure}[ptb]
\centering
\includegraphics[width=3.35in]{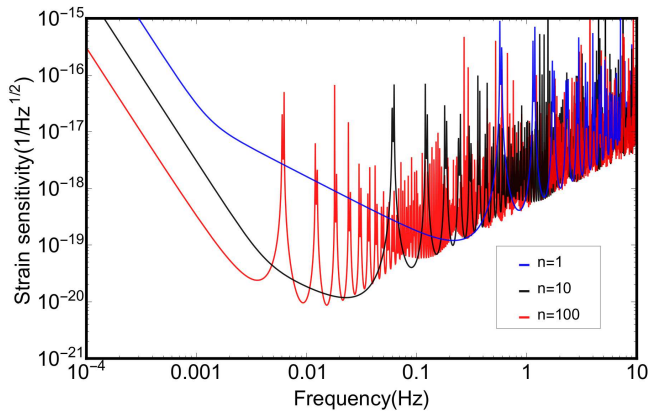} \caption{(Color online) Sensitivity
curves for $n=1$, $10$, and $100$-trip one-wave laser propagation with
acceleration noise included.}%
\label{fig5}%
\end{figure}

The sensitivity curve can be obtained by%
\begin{equation}
h_{f}=\sqrt{S_{h}\left(  f\right)  }=\sqrt{\frac{S_{p}\left(  f\right)
}{\Gamma\left(  f\right)  }},
\end{equation}
where the spectral density $S_{p}\left(  f\right)  $ includes contributions
from quantum projection noise and acceleration noise, and $S_{h}\left(
f\right)  $ is derived from the GW signal. The sensitivity curves for $n=1$,
$10$, and $100$ are shown in Fig. \ref{fig5}. Indeed, our $n$-way method is
found to be capable of improving sensitivity in the low frequency range with
its increased transfer function, but it cannot overcome the acceleration noise
floor, which leads to the non-ideal improvement. In fact, the frequency at the
point of inflection (approximately 3mHz) is determined by the quantum
projection noise that is about $\sqrt{N}\sim10^{-3}$Hz. Thus, our method
provides a way to reach the same sensitivity near the inflection point of the
frequency with decreased length between two atomic clocks but with increased
repeated number of the laser pulses, as presented in Fig. 5 where the case for
$n=100$ is equivalent to that for the distance $d=5\times10^{10}$ m.

\section{Conclusion}

In this paper, we propose a $n$-way scheme of GW detection with optical atomic
clocks. At first, we have compared the single two-way measurement or two
one-way measurements with the single one-way measurement and found that
although two one-way measurements are optimal, but it is not easy to extend to
more ways and in particular, its optical measurement cannot shift to lower
frequency. So the two-way measurement is focused, since it has an advantage
that shifts the optimal measurement to lower frequency, which is a special
case ($n=2$) in our recycling scheme. For our method suggested in fourth
section, it is found that the signal of $n$-ways summation can improve the
sensitivity for low-frequency GW by reducing the quantum projection noise over
a broad frequency range. Our method, in essence, is equivalent to the
operation of increasing the distance between atomic clocks by increasing the
number of the repeated sending pulses within the permission of other operation
conditions, i.e. atomic linewidth. This means that if we want to detect lower
GW frequencies, we don't need to set up the scheme by taking a larger distance
between atomic clocks, and it can be reached only by repeating to send some
pulses back and forth, but the repeated number is limited by the atomic
linewidth. This is different from the average over many measurements, since
the latter cannot shift the optimal measurement point to lower frequencies. We
thus conclude that our $n$-way protocol presents a nice improvement for
space-based optical interferometry with baseline length of $10^{8}$ meters,
which needs lower optical power compared to those space-based detectors with
longer arm length. We have also studied the situation\ including the
acceleration noises and found that the improvement of sensitivity would be
restrained below the inflection point of frequency which is determined by the
quantum projection noise. So a better method is required to reduce the
acceleration noises for the further improvement of sensitivity in the
low-frequency GW detection.

\section{Acknowledge}

We thank Li You for the helpful discussions and insights. This work is
supported by NSFC (No. 11654001 and No. 91636213). B. Zhang suggested and
planned the work, F. He and B. Zhang finished all analyses and calculations
together, and F. He made all the figures.

\end{document}